\documentclass[]{aa}
\usepackage{amssymb}
\usepackage{epsfig}
\usepackage{times}
\begin{document}
\thesaurus{12.04.1 - 10.08.1 - 10.19.3 - 08.12.2}
\title{Microlensing towards different Galactic targets}
\author{Lukas Grenacher\inst{1,2}, Philippe Jetzer\inst{1,2}, 
Marcus Str\"assle\inst{2} and Francesco De Paolis\inst{3}}
\offprints{lugre@physik.unizh.ch}
\institute{
Paul Scherrer Institut, Laboratory for Astrophysics,
CH-5232 Villigen PSI
\and
Institut f\"ur Theoretische Physik
           der Universit\"at Z\"urich,
           Winterthurerstrasse 190,
           CH-8057 Z\"urich, Switzerland
\and 
Department of Physics and INFN, University of Lecce, CP 193, I-73000 Lecce,
Italy}

\date{Received; accepted}

\maketitle
\markboth{Microlensing Events and Dark Objects by Mass and Time Moments}{}

\begin{abstract}
We calculate the optical depth and the number of events due to gravitational
microlensing towards the Galactic 
bulge, the spiral arm directions $\gamma$ Scutum,
$\beta$ Scutum, $\gamma$ Normae,
$\vartheta$ Muscae and some dwarf galaxies in the halo of the Galaxy.

Using the events found by the MACHO collaboration during their first
year of observation towards Baade's Window
we estimate the mass functions for the bulge and disk populations following 
the mass moment method. We find that the mass function can be described
by a decreasing power-law with slope 
$\alpha \simeq 2.0$ in both cases and a minimal mass of $\sim 0.01$ $M_{\odot}$
for the bulge and $\sim 0.02$ $M_{\odot}$ for the disk, 
respectively.
Assuming that the obtained mass function for the disk is also valid in the spiral arms,
we find that the expected number of events towards the spiral arms is in
reasonable agreement with the observations. However, the small number of 
observed events does not yet constrain much the different parameters
entering in the computation of the mass function.

To study the influence of the Magellanic 
Clouds on the shape and the velocity dispersion in the halo we perform
a N-body simulation.
We find that their presence induces a slight flattening of the halo
($q_{H}\simeq 0.8$). As a result the expected number of microlensing
events towards some targets in the halo, such as the LMC or the SMC,
decreases by about 20\%, whereas due to the the modification induced 
on the velocity dispersion the event duration increases.

\end{abstract}

\section{Introduction}

During the last ten years amazing progress was made by exploring the dark 
component of our own Galaxy by means of gravitational microlensing, as proposed by 
Paczy\'nski in 1986.
Microlensing allows the detection of MACHOs (Massive Astrophysical 
Compact Halo Objects) in the mass range $10^{-7}\lesssim M/M_{\odot}\lesssim 1$
(De R\'ujula 
et al. \protect\cite{Jet92}) located in the Galactic halo, as well as 
in the disk or bulge of our Galaxy (Pazcy\'nski \protect\cite{Pac91}, 
Griest et al. \protect\cite{Gri91}). 
Microlensing searches have nowadays become an important branch
in astrophysics, especially for the study of the structure of the Milky Way
and even for globular clusters (Jetzer et al. 1998).

Today, more than half a dozen groups are active in observing 
microlensing events towards different lines of sight, and they reported 
so far several hundreds of events, most of them towards the Galactic 
bulge (Alcock et al. \protect\cite{Alc97a}, Udalski et al. 
\protect\cite{Uda94}, Alard et al. \protect\cite{Duo97}), some towards the 
spiral arms (Alcock et al. \protect\cite{Alc97a}, Derue et al.
\cite{derue}), about 15 events 
towards 
the LMC 
(Alcock et al. \protect\cite{Mac93}, \protect\cite{Alc97b}, Auburg et al. 
\protect\cite{Ero93}) and two events towards the SMC (Alcock et al. 
\protect\cite{Alc97c}, Palanque-Delabrouille et al. \protect\cite{Pal97}, 
Udalski et al. \protect\cite{Uda97}).
The necessity to extend the microlensing target regions, as for instance
towards M31 (Crotts \& Tomaney 1996, Ansari et al. 1999), is widely recognized
in order to better characterize the MACHO distribution throughout the
Milky Way and the Galactic halo.

In this paper we calculate
the optical depth and the number of events due to gravitational
microlensing towards the bulge, the spiral arm directions $\gamma$ 
Sct, $\beta$ Sct, $\gamma$ Nor,
$\vartheta$ Mus and some dwarf galaxies in the halo of the Galaxy.
To that purpose we use the method of
mass and time moments (De R\'ujula, Jetzer and Mass\'o \cite{Jet90}).
Based on a model of our Galaxy and
taking into account the different populations of objects that can act either
as lenses or as sources in microlensing events, 
we compute the different mass and time moments, which are needed for
the determination of the various microlensing quantities.
Due to the extension of the bulge we include in the calculation
of the optical depth and the expected number of events the fact
that the number density of sources varies with distance.

The 41 events published by the MACHO collaboration towards Baade's Window
allow us to calculate the mass functions
for the objects acting as lenses towards the Galactic 
bulge. The result is described by a decreasing power-law with
$M_{min}\simeq 0.012 M_{\odot}$ for lenses in the bulge and 
$M_{min}\simeq 0.021
M_{\odot}$ for lenses in the disk and a slope
$\alpha\simeq 2.0$ in both cases.
Moreover, assuming that the disk mass function 
is the same also in the spiral arms,
we find that the calculated number of events towards the spiral arms is in
good agreement with the events reported by the EROS II collaboration.

However, the small number of 
observed events towards the spiral arms does not yet constrain much the 
different parameters of the mass function.

We emphasize that, due to the small number of events at disposal,
the mass functions we found for the bulge and the disk should be 
considered as an illustrative example of how the mass moment method
can be used to get useful information on the physical parameters. Indeed,
when the 
many new events which have been observed in the meantime will be published,
more accurate results will be obtained.

To study the influence of the Magellanic
Clouds on the shape and the velocity dispersion in the halo we perform a
N-body simulation.
We find that their presence induces a slight flattening of the halo
($q_{H}\simeq 0.8$) starting from an initially isotropic distribution. 
As a result the expected number and the duration of microlensing
events towards targets in the halo gets modified. 
  
The paper is organized as follows: in Sect. 2 we present our Galaxy model,
in Sect. 3 we introduce the method of mass and time moments. 
The calculation of the 
various microlensing quantities towards the Galactic center and the spiral 
arms is done in Sect. 4, while Sect. 5 is dedicated to different
targets in the halo of the Galaxy.
A short discussion in Sect. 6 concludes the paper. The N-body simulation
of the Galactic halo is presented in the Appendix.

\section{Galaxy models}
Our Galaxy model consists of three components: a bulge, a disk and a halo,
which we discuss in the following.

\subsection{Mass density}

Following Han and Gould (\cite{Han95}) we assume a triaxial bulge with a 
density law (Dwek et al. 1995)
\footnote{The integral $\int_{-\infty}^{\infty}\exp(-s^2/2)dx^{\prime}dy^{\prime}dz^{\prime}$
evaluates to
$6.57\pi abc$. In the literature sometimes the wrong normalization
$1/(8\pi abc)$ is used instead.}
\begin{equation}
    \rho_{B}(x^{\prime},y^{\prime},z^{\prime})=\rho_{\circ B}~e^{-s^2/2}=\frac{M_{B}}{6.57\pi abc}~e^{-s^2/2}\,,
    \label{eq:bulge_dens}
\end{equation}
with $s^4=\left(x^{\prime^2}/a^2+y^{\prime^2}/b^2\right)^2+z^{\prime^4}/c^4$.
$M_{B}\simeq 1.8\times 10^{10}~M_{\odot}$ is the estimated bulge mass. The 
length scales are: $a=1.58$ kpc, $b=0.62$ kpc and $c=0.43$ 
kpc ($x^{\prime}$ and $y^{\prime}$ are defined along and perpendicular to the bar shaped 
bulge in the Galactic plane). We take for the inclination angle between the bar major axis and the line of 
sight towards the Galactic center a value of $\Phi_{\circ}=20^{\circ}$.
The influence of this parameter on the results will be discussed.

We consider a disk model with a
thin and a thick component and a
density distribution as given by Bahcall et al. (\cite{Bah83}) 
and Gilmore et al. (\cite{Gil89})
\begin{eqnarray}
& &    \rho_{D}(R,z) = \frac{1}{2}~\exp\left[-\frac{(R-R_{0})}{h}\right]
\times \label{eq:disk_dens_2fach}
 \nonumber\\
& &               
\left[\frac{\Sigma_{\rm{thin}}}{H_{\rm{thin}}}\exp\left(-
\frac{|z|}{H_{\rm{thin}}}\right)+
\frac{\Sigma_{\rm{thick}}}{H_{\rm{thick}}}\exp\left(-
\frac{|z|}{H_{\rm{thick}}}\right)\right]\,,
\end{eqnarray}
where $R,\,z$ are cylindrical coordinates.
Here we adopt $H_{\rm{thin}}\simeq 0.3$ kpc for the thin and 
$H_{\rm{thick}}\simeq 1$ kpc for the thick disk component, 
whereas $h\simeq 3.5$ kpc for the length scale of
the two disks. Following Gates et al. (\cite{Gat95}), we take for
the local surface densities 
$\Sigma_{\rm{thin}}\simeq 25~M_{\odot}/\rm{pc}^{2}$ and 
$\Sigma_{\rm{thick}}\simeq 35~M_{\odot}/\rm{pc}^{2}$. $R_{0}=$ 8.5 kpc
is the distance of the solar system from the Galactic center.

The halo is assumed to be a self-gravitating isothermal 
sphere of an ideal gas in hydrostatic equilibrium.
A slightly flattened halo can then be described by the
density distribution
\begin{equation}  \rho_{H}(x,\,y,\,z)=\frac{\rho_{\circ H}}{q_H}~
\frac{R_{c}^2+R_{0}^2}
{R_{c}^2+\displaystyle\frac{z^2}{q_{H}^2}+x^2+y^2}\,,\label{eq:halo_dens}
\end{equation}
where $q_{H}$ is the axis oblateness ratio (Binney and Tremaine \cite{Bin87}), 
$R_{c}\simeq 5.6$ kpc is the core radius and 
$\rho_{\circ H}\simeq 7.9\times 10^{-3}~M_{\odot}~{\rm pc^{-3}}$ is the local dark mass density.

For the mass within $50\,$kpc we obtain for the bulge 
$M_{B}=1.8\times 10^{10}M_{\odot}$, disk $M_{D}=5.2\times 10^{10}M_{\odot}$ 
and halo
$M_{H}=4.3\times 10^{11}M_{\odot}$, respectively. 
Thus the total mass is 
$M_{\rm{tot}}\simeq 5\times 10^{11}M_{\odot}$ in agreement with e.g.
Kochanek (\cite{Koc95}). 
The calculated rotation curve 
agrees well with the measured values.

\subsection{Mass distribution}

The mass density does not determine the MACHO number density as a function of 
mass alone. Assuming the mass-distribution to be independent of
the position in the galaxy, the number density can be written as
\begin{equation}
    \left(\frac{dn}{d\mu}\right)_{i}d\mu=\left(\frac{dn_{\circ}}
    {d\mu}\right)_{i} \frac{\rho_i(\bf{r})}{\rho_{\circ_{i}}}~d\mu\,
    ,\label{eq:pos_ind_mf}
\end{equation}
where $\mu$ is the MACHO mass 
in solar mass units. The subscript $i$ stands for the bulge (B), the 
disk (D) or the halo (H). 
The MACHO number density per unit of mass, $dn_{\circ}/d\mu$, 
is normalized as follows:
\begin{equation}
    M_{\odot}\int_{\mu_{min}}^{\mu_{max}}\left(\frac{dn_{\circ}}{d\mu}\right)_{i}\mu
d\mu=
    \rho_{\circ_{i}}\;.\label{eq:mf_norm}
\end{equation}
The total mass distribution is just the sum over the components in
Eq.(\ref{eq:pos_ind_mf}). 
We assume a maximal mass $\mu_{max}\simeq 10\,M_{\odot}$ for the stars.
However, only the faint stars up to about $\mu_{up}\simeq 1\,M_{\odot}$
can contribute to microlensing events. For the bright stars $\mu \geq
1\,M_{\odot}$ we assume a Salpeter IMF ($dn_{0}/d\mu\propto \mu^{-2.35}$),
whereas for the lenses we will either assume all objects to have the same 
mass $\mu_{\circ_{i}}$ and, therefore, the mass
distribution is described by a delta function
\begin{equation}
    \left(\frac{dn_{\circ}}{d\mu}\right)_{i}=\frac{\rho_{\circ_{i}}}
    {M_{\odot}\mu_{\circ_{i}}}\delta(\mu-\mu_{\circ_{i}}) \label{eq:delta_mf}
\end{equation}
or we assume a power-law
\begin{equation}
\left(\frac{dn_{\circ}}{d\mu}\right)_{i}=C_{i}^1(\alpha_i)\mu^{-\alpha_i}\,.\label{eq:pl_mf}
\end{equation}
Accordingly the
factor $C_{i}^1(\alpha_i)$ in the power-law is fixed by the normalisation 
condition (for $i=B,\,D$) 
\begin{equation}
   \frac{\rho_{0_i}}{M_{\odot}}=C_i^1\int_{\mu_{min}}^{\mu_{up}}\mu^{1-\alpha}d\mu+
              C_i^2\int_{\mu_{up}}^{\mu_{max}}\mu^{1-2.35}d\mu\,.
\end{equation}
Assuming continuity for the mass function at $\mu_{up}=1\,M_{\odot}$ we get
\begin{equation}
C_{i}^1(\alpha_i)=\frac{\rho_{\circ_{i}}}{M_{\odot}}
\left\lbrace
\begin{array}{lcl}
          \left[\frac{1-\mu_{min}^{2-\alpha_i}}
{2-\alpha_i}+1.5809\right]^{-1}
                                            &  \rm{if} & \alpha_i \neq 2 \\
         \left[-\ln\mu_{min}+1.5809\right]^{-1} &  \rm{if} &  \alpha_i=2.
\end{array}\right.
\label{eq:c_mf}
\end{equation}

\subsection{Velocity distribution}

For sources and lenses located in the bulge we assume that the velocity
distribution along the various axes is Gaussian $f(v_y,v_z)$
$=f(v_y)f(v_z)$,
 where
\begin{equation}
   f(v_y)=\frac{1}{\sqrt{2\pi} \sigma_y}~\rm{exp}\left[
    -\frac{(v_y-\bar{v}_y)^2}{2\sigma_y^2}\right]
\end{equation}
and a corresponding distribution for $f(v_z)$. The mean velocities
$\bar{v}_y$ and $\bar{v}_z$ are supposed to be zero and the dispersion
velocities $\sigma_y$ and $\sigma_z$ are deduced from the tensor virial
theorem (Binney \& Tremaine \cite{Bin87}). Following Han \& Gould
(\cite{Han95}) the dispersion velocities in the coordinate system 
given by the principal axes of the bulge ellipsoid are $(\sigma_{x^{\prime}},
\sigma_{y^{\prime}},\sigma_{z^{\prime}})=(113.6,77.4,$ $66.3)$,
where the mean line of sight dispersion velocity is normalized to 110 
km s$^{-1}$. The projected velocity 
dispersions in Galactic coordinates (such that $x$ is along the line of
 sight, $x-y$ is in the Galactic plane and $z$ perpendicular to it) are 
then computed to be
$(\sigma_x,\sigma_y,\sigma_z)=(110,82.5,66.3)$. 
In the following it is convenient to write the distribution function
$f(v_y,v_z)$ introducing polar coordinates in the $v_y-v_z$ plane.
This way in polar coordinates
the transverse velocity distribution turns out to be
\begin{equation}
   f_{T,B}(\nu,\vartheta)d\nu d\vartheta=\frac{\nu}{\pi}~\rm{e}^{-\nu^2}d\nu 
        d\vartheta~.\label{eq:bulge_distr}
\end{equation}
The transverse velocity is then given by
\begin{equation}
   v_T=\nu\sqrt{2(\sigma_y^2\rm{cos}^2\vartheta+
\sigma_z^2\rm{sin}^2\vartheta)}~.
\end{equation}
$\vartheta$ is the polar angle and 
the variable $\nu$ is dimensionless.

In the halo we consider a Maxwellian velocity distribution with typical
dispersion velocity of $v_{H}=\sqrt{2}\sigma_H\simeq$ 220 km s$^{-1}$ 
(Paczy\'nski \cite{Pac91}). The transverse velocity distribution is then
given by
\begin{equation}
    f_{T,H}(v_{T})dv_{T}=\frac{2}{v_{H}^2}v_{T}\exp
    \left(\frac{-v_{T}^2}{v_{H}^2}\right)dv_{T}~.\label{eq:max_distr}
\end{equation} 

For lenses belonging to the disk, the velocities of the observer and the 
source transverse to the line of sight are ${\mathbf v_0}$ and 
${\mathbf v_{s}}$, respectively, and the
transverse velocity of the microlensing tube of radius $R_{E}$ (where the Einstein
radius $R_{E}$ will be defined in Sect.~3) at position $xD$ ($0 \leq x \leq 1$) is
${\mathbf v_{t}}(x)=(1-x){\mathbf v_0}+x{\mathbf v_{s}}$. Its
absolute value is
\begin{eqnarray}
  & &  v_{t}(x)=\\
& & \sqrt{(1-x)^2|{\mathbf v_0}|^2+x^2|{\mathbf v_{s}}|^2+2x(1-x)|
{\mathbf v_0}||{\mathbf v_{s}}|\rm{cos}~\vartheta}\;,\nonumber
\label{eq:lens_shift}
\end{eqnarray}
where ${\mathbf v_0}={\mathbf v_{\odot}} \cos l$ and 
${\mathbf v_{s}}$ are the solar
and the source velocities transverse to the line of sight and $\vartheta$ the
angle between them, whereas $l$ denotes the Galactic longitude.
 The solar velocity transverse to the Sun-Galactic centre 
line is denoted by ${\mathbf v_{\odot}}$, with $|{\mathbf v_{\odot}}|\simeq 
220~\rm{km~s}^{-1}$. 
The distribution of the transverse velocity of the lens thus is 
\begin{eqnarray}
  &&  g_{T}(v_{T},\vartheta)~dv_{T}d\vartheta=\nonumber \\
  &&  \frac{1}{\pi v_{D}^2}~v_{T}~
    \exp \left(\frac{-(\mathbf{v_{T}}-\mathbf{v_{t}})^2}{v_{D}^2}\right)
    dv_{T}d\vartheta~,
\label{eq:rel_vel}
\end{eqnarray}
with a velocity dispersion $v_{D}\simeq 30~ \rm{km~s}^{-1}$ 
in the disk (Pac\-zy\'n\-ski \cite{Pac91}). 

Due to the relative position $\gamma$ Nor and $\vartheta$ Mus are approximately
moving towards us, whereas $\gamma$ and $\beta$ Sct are moving away from us.
Hence, we will neglect the dispersion in the transverse velocity for
sources located in these particular fields towards the spiral arms.
We take into accout the dispersion velocity of the lenses in the disk using
an expression as given in Eq.(\ref{eq:max_distr}) 
but with a velocity dispersion
$v_{D}\simeq 30~ \rm{km~s}^{-1}$.

\section{Mass and time moments}

\subsection{Optical depth and differential number of events}

To estimate the microlensing probability one introduces the 
optical depth $\tau_{opt}$ defined as
\begin{equation}\label{pj3}
    \tau_{opt}=\int_{0}^{1}\frac{4\pi G}{c^2}\rho(x)D^2x(1-x)\,dx\,,
\end{equation}
with $\rho(x)$ the mass density at the distance 
$s=xD$ from the observer 
along the line of sight. $\tau_{opt}$ is the probability 
that a source is found within the Einstein radius $R_E$ of a lens, where $R_E$
is given by
\begin{equation}\label{pj2}
    R_E^2=\frac{4GMD}{c^2}x(1-x)\equiv r_E^2\mu x(1-x)\,,
\end{equation}
with $x=s/D$; $D$ and $s$ are the distances to the source and  
the lens, respectively.

The  
optical depth is independent of the mass function of the lensing objects.
To get the number of microlensing events 
one introduces the differential number of microlensing events
(De R\'ujula et al. \cite{Jet90}, Griest \cite{griest}) :
\begin{eqnarray}
  &&  dN_{ev}=\\
  &&  N_{*}t_{obs}u_{\rm{TH}}2v_{T}f_{T}(v_{T})Dr_{E}\sqrt{\mu x(1-x)}
    ~\frac{dn}{d\mu}d\mu dv_{T}dx\,,\nonumber
    \label{l3}
\end{eqnarray}
assuming an experiment that monitors $N_{*}$ stars during a period
$t_{obs}$. 
This quantity yields the number of microlensing events with a 
magnification above a certain threshold $A_{\rm{TH}}$ (where 
$A_{\rm{TH}}=(u_{\rm{TH}}^{2}+2)/(u_{\rm{TH}}(u_{\rm{TH}}^{2}+4)),~
A_{\rm{TH}}=1.34$ for $u_{\rm{TH}}=1$). 

\subsection{Method of mass moments}
A systematic way to describe the relevant quantities in a
microlensing experiment is the method of mass moments  
(De R\'ujula et al. \cite{Jet90}), which are defined as
\begin{equation}
    <\mu^{m}>=\int_{\mu_{min}}^{\mu_{up}} 
d\mu ~\epsilon_{n}(\mu)~\frac{dn_0}{d\mu}~\mu^{m} \,,
     \label{eq:mamo}
\end{equation}
with $m=(n+1)/2$ and 
\begin{equation} 
    \epsilon_{n}(\mu)\equiv \frac{\int dN_{ev}^{*}(\bar\mu)~
\epsilon (T)~\tau ^{n}}{\int dN_{ev}^{*}(\bar\mu)~\tau ^{n}}~.
    \label{eq:eps}
\end{equation}
The sampling efficiency $\epsilon(T)$ is given by the experiment and 
$\tau=(v_{i}/r_{E})T$, where $T$ is the time scale of the observed event
(for which we adopt the definition that it is the time needed
to cross the Einstein radius)
and $v_i$ stands for the dispersion velocity in the bulge, disk or halo.
$\epsilon_{n}(\mu)$ is called efficiency function and 
measures the fraction of 
the total number of microlensing events that, for a fixed MACHO mass 
$M=\mu M_{\odot}$, meet the condition 
$T_{min} \leqslant T \leqslant T_{max}$.
$dN_{ev}^{*}(\bar\mu)$ is defined as 
$dN_{ev}$ with, however,  
$dn_{0} /d\mu$ assumed to be a delta function $\delta 
(\mu -\bar{\mu})$.
In other words, $\epsilon_{0}(\mu)$ indicates 
how efficient the experiment is to detect a MACHO with a given mass 
$M=\mu M_{\odot}$.

$<\mu^{m}>$ is related to the cumulative $n^{th}$ moment of $\tau$ 
constructed from the observations as follows
\begin{equation}
    <\tau^{n}>=\sum_{{\rm events}}\tau^{n}\,.
\end{equation}
Due to the insertion of $\epsilon_{n}(\mu)$, 
the theoretical expression for the time moment
\begin{equation}
    <\tau^{n}>=\int dN_{ev}\epsilon_{n}(\mu)\tau^{n}
\label{20}
\end{equation}
factorises as follows
\begin{equation}
    <\tau^{n}>=Vu_{TH}\gamma(m)<\mu^{m}>~,\label{eq:tau_mo}
\end{equation}
with
\begin{equation}
    V\equiv 2N_{*}t_{obs}Dr_{E}v_{i}\,.
\end{equation}
$\gamma(m)$ is a quantity defined by the model, which depends on the spatial 
and the velocity distributions.

The mass moment $<\mu^{m}>$ is related to $<\tau^{n}>$, as given from the 
measured values of $T$, by
\begin{equation}
    <\mu^{m}>=\frac{<\tau^{n}>}{Vu_{TH}\gamma(m)}\,. 
    \label{eq:mamo_gamma} 
\end{equation}
Some moments are directly related to physical quantities:
the number of events is $N_{ev}^{\rm{obs}}=<\tau^0>$.
The mean local density of MACHOs (number per pc$^3$) is $<\mu^0>$ 
and the average local mass density in MACHOs (solar masses per pc$^3$) is 
$<\mu^1>$. Thus the mean MACHO mass is given by
\begin{equation}
    \frac{<\mu^1>}{<\mu^0>}=\frac{<\tau^1>}{<\tau^{-1}>}~\frac{\gamma(0)}
{\gamma(1)}~,\label{eq:meanmass}
\end{equation}
in units of solar masses, where $<\tau^1>$ and $<\tau^{-1}>$ are 
determined through the observed microlensing events.
The average event duration $<T>$ can be expressed as follows
\begin{equation}
   <T>=\frac{r_{E}}{v_{i}}\frac{<\mu^1>}{<\mu^{1/2}>}\frac{\gamma(1)}
    {\gamma(1/2)}~,
\label{eq:mamo_t}
\end{equation}
where again $i$ stands for the different components of the Galactic model.

\subsection{Sources in the Galactic centre}

Due to the extension of the bulge one has to take into account
that the number density of possible source stars varies
with distance $D$. This effect has to be taken into account
in the calculation of the optical depth and the number of 
microlensing events.
Following Kiraga \& Paczy\'nski (\cite{Kir94}), the volume of space
varies with distance as $D^{2}dD$ and the number of detectable stars
as $D^{2 \beta}$, assuming that the fraction of stars brighter
then some luminosity $L$ is proportional to $L^{\beta}$, where $\beta$ is a 
constant. In the following we will use a power-law luminosity function
$\sim L^{-1}$ (i.e. $\beta=-1$), which is appropriate for main sequence
source stars.
An observable quantity towards the bulge, which we shall generically denote
as $g_{B}$, is then obtained as
\begin{equation}
    g_{B}=\frac{\int_{0}^{\infty}\varphi(D)g(D)dD}
          {\int_{0}^{\infty}\varphi(D)dD}\;,
      \label{eq:source_distr}
\end{equation}
with $\varphi(D)=\rho_{B}D^{2+2\beta}$ and $g(D)$ an observable quantity as
for instance  $\tau(D)$ or $N_{ev}(D)$. $\rho_{B}$
is the number density of stars as given in Eq.(\ref{eq:bulge_dens}) and
also varies as a function of $D$. 

Accordingly, Eq.(\ref{20}) and the subsequent relations which define the 
moments get modified, due to the additional integration
over the distance $D$. It is, therefore, necessary
to use the time $T$ instead of $\tau$ (which via $r_{E}$ depends on $D$).
Otherwise, however, the calculation goes through similarly and the 
factorization
of the mass moments is again achieved. Indeed, the $D$ 
dependence in the efficiencies $\epsilon_n(\mu)$ is negligible as we 
checked numerically.

Eq.(\ref{20}) then becomes
\begin{equation}
     <T^{n}>=\tilde{V}(n)\tilde{\gamma}(m)<\mu^{m}> \label{eq:T_mo}
\end{equation}
with
\begin{eqnarray}
     \tilde{V}(n)&=&2 N_{*}t_{obs} u_{TH}\left[\frac{2}{c}
                 \sqrt{G M_{\odot}}\right]^{n+1}v_{i}^{1-n}\nonumber \\
                 &=&2\left(4.4 \times 10^{-7}\right) ^{n+1}\times 10^{6n-6}
                   \times \nonumber \\
               && N_{*} u_{TH} \left( \frac{t_{obs}}{\small{1~{\rm year}}}\right)
               \left( \frac{v_{i}}{\small{{\rm km~s^{-1}}}}\right)^{1-n} \\
 {\rm and} ~~\tilde{\gamma}(m)&=&\frac{1}{N_{B}}\int \rho_B D^{m+3+2\beta}
\gamma(m)dD~,\label{35} 
\end{eqnarray}
where $N_{B}=\int_{0}^{\infty}\varphi(D)dD=\int_0^{\infty}
\rho_{B}D^{2+2\beta}dD$.

On the other hand $<T^{n}>$ is determined from the observations by
$<T^{n}>=\sum T^{n}$ ($T$ given in days).  
Thus, the mean mass towards the bulge turns out to be
\begin{eqnarray}
      \frac{<\mu^{1}>}{<\mu^{0}>}&=&\frac{(v_{i}c)^{2}}{4G M_{\odot}}
        \frac{<T^{1}>}{<T^{-1}>}\frac{\tilde{\gamma}(0)}{\tilde{\gamma}(1)}
        \nonumber \\
                 &=&5.35\left(\frac{v_{i}}{\small{{\rm
               km~s^{-1}}}}
        \right)^{2} \frac{<T^{1}>}{<T^{-1}>}
                  \frac{\tilde{\gamma}(0)}{\tilde{\gamma}(1)}
\end{eqnarray}
and the event duration is
\begin{eqnarray}
        <T>&=& \frac{2\sqrt{G M_{\odot}}}{v_{i}c}
            \frac{\tilde{\gamma}(1)}{\tilde{\gamma}(1/2)}
            \frac{<\mu^{1}>}{<\mu^{1/2}>} \\
            \label{eq:dur_bulge}
          & = &158\left(\frac{v_{i}}{\small{{\rm km~s^{-1}}}}\right)^{-1} \frac{\tilde{\gamma}(1)}{\tilde{\gamma}(1/2)}
            \frac{<\mu^{1}>}{<\mu^{1/2}>}~~{\rm days}~.\nonumber
\end{eqnarray}

We now turn to the computation of the $\gamma(m)$,
which enter into Eq.(\ref{35}).
For microlensing events towards the bulge, we have to distinguish between
two cases: a) both the source and the lens are located in the bulge 
(bulge-bulge event), b) the source belongs to the bulge and the lens to 
the disk (bulge-disk event).
We suppose the velocity distribution of sources and lenses in the bulge
to be Maxwellian as given by Eq.(\ref{eq:max_distr}).
In this case the relative motion between observer and source can
be neglected and the $\gamma (m)$-function as defined in Eq.(\ref{eq:tau_mo})
through the mass moments is given by
\begin{equation}
    \gamma_{BB}(m)=\Xi(2-m)\hat{H}(m)~,\label{eq:gammab1}
\end{equation}
where
\begin{eqnarray}
    &&\hat{H}(m)=\int_0^{1}\left(x(1-x)\right)^{m}\frac{\rho_{B}(x)}
{\rho_{\circ_{B}}}dx\\ \label{eq:gammab2}
   &&\Xi(2-m)\equiv \int_0^{\infty}d\nu \int_0^{2\pi}d\vartheta
     \left(\frac{v_T\left(\nu\right)}{v_B}\right)^{1-n}f_{T,B}(\nu,\vartheta)~.\label{eq:gammab3}
\end{eqnarray}
For lenses belonging to the disk, the transverse velocity 
distribution is given by Eq.(\ref{eq:rel_vel}), 
and we get 
\begin{eqnarray} 
    \gamma_{BD}(m)&=&2~\int_0^{2\pi}d\vartheta \int_0^{\infty}dv_{s}
f_{T,B}(v_{s},\vartheta) \nonumber \\
   &\times& \int_0^1dx~\frac{\rho_{D}(x)}{\rho_{\circ_{D}}}~[x(1-x)]^{m}~
      \rm{e}^{-\eta ^2} \nonumber \\
   &\times& \int_0^{\infty}dy~y^{3-2m}I_0(2\eta y)~
        e^{-y^2}\,.
       \label{eq:gamma_gc}
\end{eqnarray}
with $y=v_{T}/v_{D}$, and $\eta=v_{t}/v_{D}$. The index $D$ labels the disk, 
$v_{D}\approx 30~\rm{km~s}^{-1}$ is the dispersion velocity in the disk and 
$I_0$ is the modified Bessel function of order 0 defined as
\begin{equation}
    I_0(x)=\frac{1}{\pi}\int_0^{\pi}e^{x~\rm{cos}(\beta)}~d\beta~.
\end{equation}
The source stars belong to the bulge and are supposed to follow the 
distribution $f_{T,B}(v_{s},\vartheta)$ as given in Eq.(\ref{eq:bulge_distr}).

\subsection{Sources in the spiral arms and in the halo}
If the source as well as the lens is located in the disk, the small relative 
transverse velocity between them can be neglected, as explained above
and we can assume a Maxwellian velocity distribution as given in Eq.
(\ref{eq:max_distr}). We can perform the $\vartheta$ integration and the 
corresponding $\gamma (m)$-function turns out to be as in 
Eq.(\ref{eq:gammab1}), with Eq.(\ref{eq:gammab3}) replaced by
\begin{eqnarray}
    \Xi(2-m)&\equiv& \int_0^{\infty}\left(\frac{v_{T}}{v_{D}}\right)
         ^{1-n}f_{T,H}(v_{T})dv_{T}\nonumber \\
        &=&\Gamma(2-m)~,\label{eq:gammad}
\end{eqnarray}
where $v_{D}\simeq 30~\rm{km~s}^{-1}$ is the dispersion velocity in the disk.
Sources in the halo, like LMC, SMC or dwarf galaxies are more distant, and 
thus the relative motion between observer and source can also be neglected in 
leading order.
Assuming in addition a Maxwellian velocity distribution
the resulting $\gamma (m)$-function turns out to be as in Eqs.
(\ref{eq:gammab1}) and (\ref{eq:gammad}), with $v_D$ replaced by $v_H$, the 
dispersion velocity in the halo.

\section{Lensing towards the Galactic bulge and some spiral arm regions}

Let us apply the formalism presented above to different
lines of sight towards the bulge and some
fields in the spiral arms. In the bulge we focus 
on the direction of Baade's Window ($l=1^{\circ}$, $b=-3.9^{\circ}$) and
in the spiral arms towards 
$\gamma$ Nor ($l=-28.8^{\circ}$, $b=-2.7^{\circ}$), where some 
microlensing events have already been observed. 

\subsection{Optical depth}

\begin{figure}
\leavevmode
    \epsfig{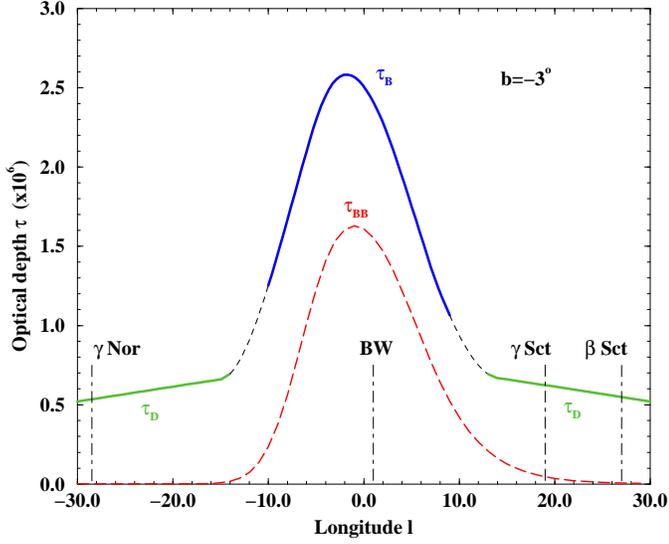}
    \caption{\small{The optical depth $\tau$ 
(in units of $10^{-6}$) for the 
different components of the Galaxy as a function of the Galactic
longitude $l$ for a fixed
latitude $b=-3^{\circ}$. For the disk contribution ($\tau_{D}$)
the distance is fixed at 8 kpc,
while for the bulge contribution ($\tau_{B}$) we integrated over the  
distance distribution of the lenses and sources. $\tau_{BB}$ denotes the
fraction of $\tau$ for  
source and lens located in the bulge (where the tilt angle 
$\Phi_0$ is taken to be $20^{\circ}$). The dotted lines are just an
interpolation between the bulge and the disk regions.
The approximate positions of $\gamma$ Nor, $\gamma$ and $\beta$ Sct and
Baade's Window (BW) are indicated.}}
    \label{fig:longbw}
\end{figure}

In Fig. \ref{fig:longbw} the optical depth $\tau$ is shown as a function of 
the Galactic longitude $l$ 
for a fixed value of the latitude $b=-3^{\circ}$. The approximate position
of $\gamma$ Nor, $\gamma$ and $\beta$ Sct and Baade's Window are
indicated. The idea is just to give an impression of the global behaviour
of the optical depth towards Baade's Window and the spiral arms. The 
values towards these targets are given below.

The dashed line $\tau_{BB}$ is the contribution due to lenses and
sources located in the bulge for which we have taken into account
the varying distances. One clearly sees that it is the major
contribution to the optical depth in the range
$-8^{\circ}\lesssim l \lesssim 10^{\circ}$. 
The tilt angle $\Phi_{\circ}$ (for which we assume $\Phi_{\circ}=20^{\circ}$)
of the bar major axis with respect to the 
Sun-Galactic centre line leads towards the bulge to a slight
asymmetry in the optical depth with respect to $l=0^{\circ}$. 
 
The curve labeled by $\tau_{B}$ is the total optical depth, 
which contains also the contribution
due to lenses located in the disk and the source star in the bulge for which
we again take into account the varying distances.

Towards Baade's Window ($l=1^{\circ}$, $b=-3.9^{\circ}$) 
the total optical 
depth $\tau_B=1.9\times 10^{-6}$
is the sum of the 
disk $\tau_{D}=0.7\times 10^{-6}$ 
and the bulge $\tau_{BB}=1.2\times 10^{-6}$
contributions.
The bulge contribution is about 
1.7 times more important than the disk. 
This value is in reasonable 
agreement with the optical depth found by the MACHO 
group: $\tau_{B}=(2.4\pm 0.5)\times 10^{-6}$ 
(Alcock et al. \cite{Alc97a}). The OGLE collaboration found a somewhat higher 
value: 
$\tau_{B}=(3.3\pm 1.2)\times 10^{-6}$ (Udalsky et al. \cite{Uda94}). 
We notice that
it has  been pointed out that the contribution of unresolved
stars might be quite significant and as a result imply that the 
measured optical depth is overestimated (Alard \cite{Alard}).
If this is the case the measured value has to be regarded as 
an upper limit to the true value.

By varying the tilt angle from $15^{\circ}$ to $30^{\circ}$ the optical
depth $\tau_{BB}$ for the bulge decreases from $1.33\times 10^{-6}$ to
$0.97\times 10^{-6}$ (and the total
optical depth from $2.02\times 10^{-6}$ to $1.68\times 10^{-6}$).

Outside the bulge region both the source and the lens belong
to the disk population and the total optical depth is labelled in Fig.
\ref{fig:longbw}
by $\tau_{D}$. In this case we assume all source stars to
be located at the same distance of 8 kpc.

Observations towards 27 different fields in the spiral arms are 
included in the scientific program of the EROS II (Derue et al. \cite{derue})
and the MACHO teams. 
The 27 fields observed by EROS II
belong to 4 dense regions of the spiral arms: 
$\gamma$ Sct, $\beta$ Sct, $\gamma$ Nor and $\vartheta$ Mus. 
The star 
density lies between 400\,000 and 700\,000 stars per field, 
leading to about 10 millions observable stars 
(Mansoux \cite{Man97}).
Their average Galactic coordinates, distances, observed number of 
stars, number of events found so far 
during an observation time of 1.7 years and 
the calculated optical depth are reported in Table \ref{tab:fields_sa}.

EROS II found altogether 3 events (see Table \ref{tab:fields_sa})
and estimated an optical depth, averaged over the four
directions, of $\tau = 0.38^{+0.53}_{-0.15}
\times 10^{-6}$ (Derue et al. \cite{derue}).
For comparison we get an averaged optical depth of $\tau = 0.42\times 10^{-6}$
using our model, which is in good agreement with the EROS II value.

\subsection{The mass functions}

Towards the bulge we compute
the mass functions using the first year 
data of the
MACHO collaboration taken in  1994. Observing during 
190 days star fields with a total of 12.6 million
source stars towards Baade's Window, they found 40 microlensing events
(neglecting the double lens event). 

Due to the limited amount of data, we make the Ansatz 
$dn_{0}/d\mu \sim C_i^1(\alpha_i)\mu ^{-\alpha_i}$ and use then the mass moments
to determine the parameters $\alpha$ and $\mu_{min}$.
$C_i^1(\alpha_i)$ is defined by
Eq.(\ref{eq:c_mf}) and we assume a maximal mass ($\mu_{max}$)
for the stars of $\sim 10 M_{\odot}$.
The upper limit $\mu_{up}$ of the integration in Eq.(\ref{eq:mamo})
is set equal to 1 $M_{\odot}$, since more massive and thus brighter 
stars will not contribute much to microlensing events. 
To evaluate Eq.(\ref{eq:mamo}) we first have to calculate 
the efficiency functions defined by Eq.(\ref{eq:eps}) using the 
sampling efficiency $\epsilon(T)$ as given by the MACHO team
for their first year events towards the Galactic bulge. 

We notice that one should also take into account the
contribution due to unresolved stars, which for the bulge might
be relevant and induces changes in the optical depth and the microlensing rate
(Alard \cite{Alard}; Han \cite{Han}).
Indeed, if a faint unresolved star is lensed close enough to a resolved star,
the event will be seen by the microlensing experiment and attributed
to the brighter star. The blending biases the event towards a shorter 
duration, leading to an overestimate of the amount of these events.
As a result the mass function will be shifted towards lower 
values of $\mu_{min}$. 
A quantitative estimate of the induced error on our results
is, however, difficult, since it would require a detailed knowledge
of the data analysis and the determination of the sampling efficiency.
Hopefully, this information will be available when the new
microlensing observations towards the bulge will be made available
by the observational teams.

The mass 
moments as computed by Eq.(\ref{eq:mamo}) 
are compared to the corresponding time moments derived from the observed
microlensing events through the expression given in 
Eq.(\ref{eq:mamo_gamma}). 
Thus with the mass moments, 
$<\mu^{0}>$ and $<\mu^{1/2}>$, we get two
equations, which then are solved with respect
to the unknown quantities $\alpha$ and $\mu_{min}$.
Further moments can be used for instance to check whether the 
values we obtain for $\alpha$ and $\mu_{min}$ are consistent.
Since we do not know the fraction of the observed events due to lenses 
located in the bulge, respectively in the disk, we assume that this fraction
is equal to the corresponding ratio of the optical depths
hence $\tau_{BB}/\tau_{B}\simeq N_B/N_{tot}(\simeq 25.1/40)$. This assumption is well
fulfilled if the mean event duration of the bulge and disk events are 
the same. We will then a posteriori verify if this condition is
matched by the results we get and thus see whether our assumption
is consistent.

With the above procedure we get for the parameters of the power-law
mass function of the lenses in the bulge the values 
$\alpha=2.0$ and $\mu_{min}=0.012$ with $C_B^1=0.34~M_{\odot}~ {\rm
pc^{-3}}$. 
Similarly we can use the disk events to determine the mass function of the 
disk population to find again $\alpha=2.0$ and instead a minimal mass of 
0.021 $M_{\odot}$ with $C_D^1=9.86\times 10^{-3}\,M_{\odot}~ {\rm
pc^{-3}}$.
The average durations we find are $\left\langle T_B\right\rangle=20.5$\,days
and $\left\langle T_D\right\rangle=19$\,days, respectively. Therefore we see
that the assumption $\left\langle T_B\right\rangle$ $\simeq\left\langle
T_D\right\rangle$ is quite well verified.

Varying the tilt angle $\Phi_{\circ}$ between $15^{\circ}$ and $30^{\circ}$
the slope $\alpha$ for the bulge mass function slightly changes 
from 1.9 to 2.15 and the minimal mass $\mu_{min}$ from 0.009 to 0.023 
whereas the number of bulge events decreases from 25 down to 22.7. 
The mean event duration for the bulge $\left\langle T_B\right\rangle$
changes accordingly from 22.1 to 19.1 days. Similar small changes
occur in the corresponding numbers of the disk events. We thus see that our 
results are robust for reasonable values of the tilt angle.  

Due to the above considerations on the unresolved stars,
the values for $\mu_{min}$ we find have to be considered as a lower limit 
on the true values. We notice that a change in $\mu_{min}$ affects also
the value of $\alpha$.
Similar studies have also been performed with different methods
by Han \& Gould (1995) and Peale (1998,1999). They find somewhat
different mass functions for the bulge, but again given the small amount of data
at disposal the errors are still large and so no definite conclusions
can be drawn. Recently, Reid et al. (\cite{reid}) using observations
from the Deep Near-Infrared Survey (DENIS) and the 2 Micron
All-Sky Survey (2MASS) have suggested
the presence of a local
unknown population of free brown dwarfs with a mass function with minimal
value as low as 0.01 $M_{\odot}$.  If confirmed it may well be plausible
to find this population also in the bulge.

\subsection{Mass and time moments towards the bulge}

For the calculation of the mass and time moments towards the bulge we have 
to distinguish between the two cases, whether the lens is located in the 
bulge or in the disk, and to choose accordingly 
the appropriate velocity distribution as given by 
Eq.(\ref{eq:bulge_distr}) or Eq.(\ref{eq:rel_vel}). 
The number of 
events and the event duration are defined by 
$N_{ev}^{obs}=<T^0>$ and Eq.(\ref{eq:dur_bulge}), respectively, 
where the moments of $T$ are given by 
Eq.(\ref{eq:T_mo}).

\begin{figure}
\leavevmode
    \epsfig{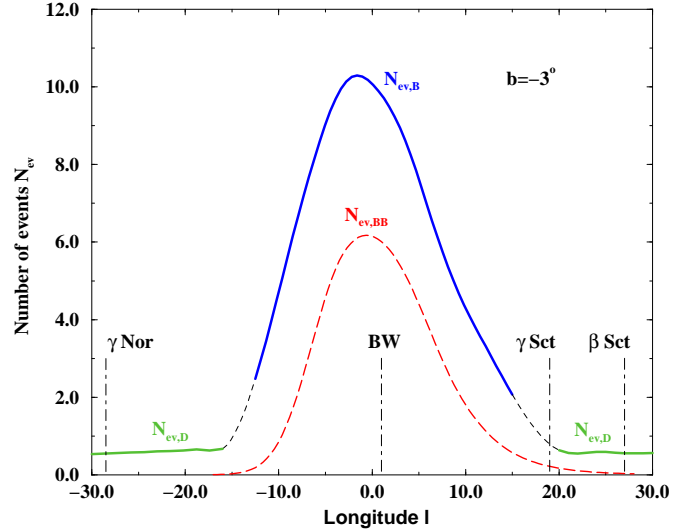}
    \caption{\small{The number of events for a delta
mass distribution with $\mu_0=1$ as a function of the longitude
with fixed latitude $b=-3^{\circ}$. The 
total exposure is assumed to be $10^6$ star-years.
The $N_{ev,BB}$ curve (source and lens in the bulge) 
reflects the position of the bulge (tilted by 
$\Phi_0=20^{\circ}$), which is important in the range 
$-8^{\circ}\lesssim l \lesssim 10^{\circ}$. In this range, the total 
number (denoted by $N_{ev,B}$)
is dominated by the events with the lens located in the bulge. 
Outside 
of this range the events $N_{ev,D}$ are due to sources located 
in the spiral arms and lenses in
the disk population. The efficiency function $\epsilon_{0} (\mu)$ is 
assumed to be 1 in this plot.}}
    \label{fig:nev1}
\end{figure}
\begin{figure}
\leavevmode
    \epsfig{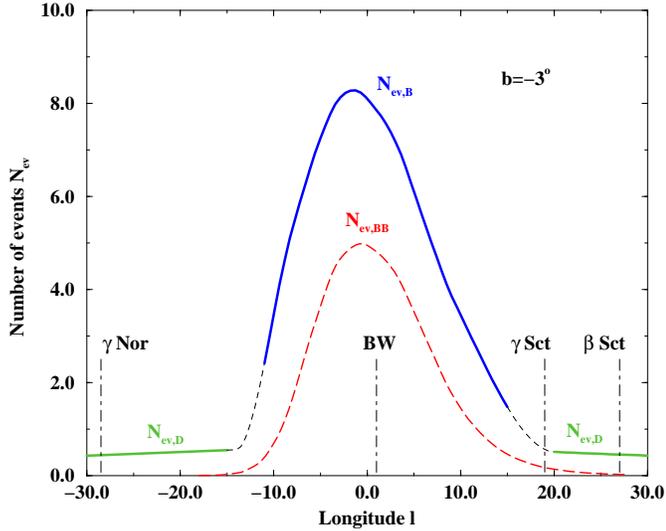}
    \caption{\small{The number of events for 
power-law mass functions with $\alpha=2.0$ and $\mu_{min,B}=0.012$ for the bulge
and $\mu_{min,D}=0.021$ for the disk population, respectively. The efficiency 
function of the MACHO collaboration is taken into account in the 
calculation to be able to directly compare with the experiment. 
The total exposure is taken to be $10^{6}$ star-years.}}
    \label{fig:nev2}
\end{figure}
Figure \ref{fig:nev1} shows the number of expected microlensing
events as a function 
of the longitude for a delta mass distribution with $\mu_0=1$,
assuming to monitor  $10^6$ stars during an observation time of 1 year.
The efficiency is taken to be 1.
Again, as in the plot of the optical depth, it shows the global behaviour
of the expected number of events towards the bulge and the spiral arm
regions.
The distance towards the spiral arms
is kept fixed at 8 kpc and the latitude at $b=-3^{\circ}$. 

$N_{ev,B}$ is the total number including also the contribution due to 
lenses located in the disk and 
sources in the bulge. For the latter we take into account
the dependence on the distance as illustrated 
in Eq.(27).
For the number of events towards the spiral arms, 
$N_{ev,D}$, the source is located in the spiral arms and the lenses 
belong to the disk population.

In Fig. 4 we plot the efficiency functions which are derived from
the sampling efficiency $\epsilon(T)$ as given by the MACHO team
towards Baade's Window.
We recall that for instance $N_{ev}$ scales like 
$\sim v~\mu^{-1/2}$, whereas 
$<T>$ as $v^{-1}~\mu^{1/2}$.
$N_{ev,BB}$ is the
contribution due to the bulge (i.e. events with source and lens 
in the bulge).

\begin{figure}
\leavevmode
    \epsfig{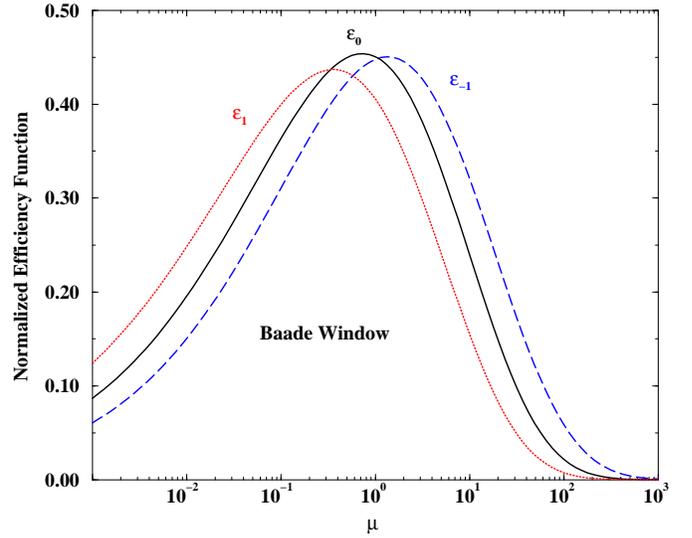}
    \caption{\small{Efficiency functions $\epsilon_{n}(\mu)~(n=-1,0,1)$ 
derived from the sampling efficiency towards Baade's Window
of the MACHO collaboration.}}
    \label{fig:epsb}
\end{figure}
 
Figure \ref{fig:nev2} shows the expected
number of microlensing events assuming the power-law
mass functions we found with the parameters 
as mentioned in Sect. 4.2.
For this plot the same efficiency $\epsilon(T)$
as given by the MACHO team towards Baade's Window 
is assumed for all directions.
All the other parameters are the same as the ones
used in Fig. \ref{fig:nev1}.

Using the values for the various parameters as given in Sect. 2 
and power-law mass 
functions with $\alpha=2.0$, $\mu_{min,B}=0.012$ for the bulge and 
$\mu_{min,D}=0.021$ for the disk population, respectively, the number
of events towards Baade's Window are consistently reproduced to be 
$N_{ev}$= 40.5. This is just the sum of $N_{ev,BB}=25.0$ and
$N_{ev,D}=15.5$, which we get when considering
12.6 million source stars  observed
during 190 days as in the MACHO first year data. The mean event 
duration, as given by Eq.(\ref{eq:dur_bulge}), is $<T>_{B}=20.5$ days for  
bulge events and $<T>_{D}=19.0$ days for disk events.
These durations have to be compared with the actual observations of
the reported 40 events with $<T_{obs}>=19.9$ days.
We thus see that the values for the durations we get using our model are in
good agreement with the observed one.
For the mean mass we get 0.091 $M_{\odot}$ for the lenses in the bulge
and 0.124 $M_{\odot}$ for the lenses in the disk, respectively.
Due to the sampling efficiency this value is actually
an upper limit of the true value (when not taking into account
the problem of the unresolved sources).
Just for comparison the theoretical value we would get for the
average mass in the bulge assuming the power-law mass function as mentioned
in Sect. 4.2, with $\epsilon = 1$,
is 0.054 $M_{\odot}$.
As expected, this value is somewhat
lower than the one inferred from the data.

\begin{table*}[h!tbp]
 \renewcommand{\arraystretch}{1.2}
  \centering
    \begin{tabular}{|l|c|c|c|c|c|c|c|c|c|c|c|c|}\hline
    &  $<l>$  &  $<b>$  &  $D$      &  $N_{*}$                &  
 $N_{oe}$   &  T          &  $\tau_{tot} $     &  $N_{ev}$  &  $<T>$&
 $\gamma (0)$&$\gamma (1/2)$&$\gamma (1)$ \\ 
    &         &         &\small{kpc}&\small{$(\times 10^{6})$}&  
   &\small{(days)}&\small{$(\times 10^{6})$}&      &\small{(days)}&&& \\ \hline
   $\gamma$ Sct &18.6   &-2.6 &6.5&1.70&1& 73& 0.41& 0.60 & 76.9&1.62&0.563&0.269   \\ \hline
   $\beta$ Sct &27 &-2.5 &6.5&1.96&0& -& 0.38    & 0.63 &77.4&1.47&0.516&0.248   \\ \hline
  $\gamma$ Nor &-28.8&-2.7   &8.0&3.01&2&98\&70&0.56   &1.29 &86.5&1.42&0.504&0.244 \\ \hline
   $\vartheta$ Mus & -53.6  & -1.8 &7.0? &1.77&0&-& 0.32&0.47&80.9&1.07&0.379&0.184 \\ \hline
    \end{tabular}
    \caption{\small{The distance $D$, the number of observed 
stars $N_{*}$ and the number of observed events $N_{oe}$ with 
their duration $T$ are given for $\gamma$ Sct, 
$\beta$ Sct, $\gamma$ Nor and $\vartheta$ Mus 
(Derue et al. \cite{derue}). The optical depth $\tau_{tot}$, 
the number of events $N_{ev}$ and 
the mean event duration $<T>$ are computed adopting 
the model outlined in Sect. 2 and a power-law mass distribution
with $\alpha=2.0$ and $\mu_{min}=0.021$. The disk
dispersion velocity is $v_D=30$ km s$^{-1}$. We assume to observe $N_{*}$
stars during an observation time of 1.7 years with a sampling efficiency as 
given by the EROS II team.}}
    \label{tab:fields_sa}
\end{table*}

\begin{table*}[h!tbp]
 \renewcommand{\arraystretch}{1.2}
  \centering
    \begin{tabular}{|l|c|c|c|c|c|c|c|}\hline
    &$\tau_{tot} $     &  $N_{ev}$  &  $<T>$&$\bar{\mu}$&
 $\tilde{\gamma} (0)$&$\tilde{\gamma} (1/2)$&$\tilde{\gamma} (1)$ \\ 
    &\small{$(\times 10^{6})$}&      &\small{(days)}&&&& \\ \hline
   Bulge& 1.20& 25.0 & 20.5&0.091&$3.82 \times 10^{2}$&$1.59 \times 10^{4}$&
$9.58 \times 10^{5}$   
     \\ \hline
   Disk& 0.70   & 15.5 &19.0&0.124&$2.83 \times 10^{5}$&$2.02 \times 10^{6}$&
$1.94 \times 10^{7}$   
    \\ \hline
   Total& 1.90   & 40.5 &19.9&-&-&-&-   \\ \hline
    \end{tabular}
    \caption{\small{The optical depth, the number of events and the mean event 
duration towards Baade's Window are computed following the model outlined 
in the text. For 
the computation of $N_{ev}$ and $<T>$ we inserted the computed power-law mass 
functions with $\alpha=2.0$ and $\mu_{min}=0.012$ for the bulge and 
$\mu_{min}=0.021$ for the disk. We assume an observation time
of 0.52 years and $12.6 \times 10^{6}$ stars.}}
    \label{tab:bw}
\end{table*}

\subsection{Mass and time moments towards the spiral arms}

The corresponding results for the various directions in the spiral arms
are summarized in Table 
\ref{tab:fields_sa}. To calculate the number of expected events
we assume the power-law mass function obtained for the disk and 
the number of observed stars (indicated in Table \ref{tab:fields_sa})
and the observation time of 1.7 years as given by the EROS II
experiment. Moreover, we adopted the sampling efficiencies quoted by
EROS II (Derue et al. 1999).\\

For the $\gamma$ Nor direction we studied how the
results change by varying the various parameters, which enter the
calculation of the number of events and their duration.
As a first point we notice that
changing the maximal mass $\mu_{max}$ of the source stars or the 
maximal mass $\mu_{up}$ of the lenses in 
the power-law model has a very minor influence on the subsequent
results and can, therefore, be neglected. 
By varying the minimal mass
$\mu_{min}$ between 0.005 and 0.05 keeping all other
parameters fixed, $N_{ev}$ decreases from 1.7 to  1.0  
and the mean event duration $<T>$ increases from 63 to 106 days. 
If the slope value
$\alpha$ of the power-law varies between 1.7 and 2.5 (for
$\mu_{min}=0.021$), the number
of events increases from 0.95 to 1.9 events with a corresponding
duration decrease from 101 to 69 days.

If we assume
instead of a power-law a delta mass distribution
with most probable masses $\mu_{0D}=0.124$
(or $\mu_{0B}=0.091$)
we get 3.1 (3.5) events with a duration of 87 (76) days.

Another important parameter is the dispersion velocity in the disk. Changing
its value between 20 and 40 km s$^{-1}$, $N_{ev}$ increases from 0.86 to 1.7 
and $<T>$ decreases from 130 to 65 days, respectively.

We thus see that the small number of observed events
does not yet constrain very much the different parameters
entering the model.
Nevertheless, once more data will be available
the allowed range for the parameters will be narrowed considerably 
and in this respect the method of mass moments will be very useful
leading to a much better knowledge of the 
structure of the bulge and the disk of our 
Galaxy.

\section{Lensing towards halo targets}

Two important unknowns in Galactic models are the shape of
the halo and the velocity distribution of the halo objects. Although it is
widely believed that dark matter halos are not spherical,
a systematic explanation how to generate deviations from spherical symmetry
is not yet available. To investigate how the presence of
massive Galactic companions such as the Magellanic Clouds
may induce a flattening of the halo, and to test how such a
presence could influence the isotropy of the dispersion velocity 
(Holder \& Widrow, \cite{Hol96}) we performed
a N-body simulation which yields, as a result,
the density and velocity distribution of the halo objects. 
For the details we refer to the Appendix. Of course our results have to be taken
as an illustration, given also the crude approximations we use.
For the targets
towards different halo directions we will adopt the flattening parameter $q_H$
and the dispersion velocities as obtained from the numerical simulation. 
These values are tabulated in Table~\ref{tab:dwarf}.

\subsection{Magellanic Clouds}

The optical depth towards the LMC for the spherical halo
model is $\tau_{{\rm LMC}}\simeq 5.33\times 
10^{-7}$ ($4.93 \times 10^{-7}$ due to the halo and $0.40\times 10^{-7}$
due to the disk contribution), 
where we used the standard parameters for the LMC
($l=280.5^{\circ}$, $b=-32.9^{\circ}$, $D=50\,{\rm kpc}$). For the halo  
core-radius we adopt $R_C = 5.6\,{\rm kpc}$. The measured value, as reported 
by the MACHO collaboration, is: 
$\tau_{{\rm MACHO}}\simeq 2.4^{+1.4}_{-0.9}\times 10^{-7}$, which 
corresponds to 
about 50\% of the above predicted value for a standard 
spherical halo (Alcock et al. \cite{Alc97a}).
Similarly, towards the SMC ($l$=302.8, $b$=-44.3, $D$=63\,kpc)
the calculated optical depth is 
$\tau_{SMC}\simeq 7.39 \times 10^{-7}$ ($7.08 \times 10^{-7}$ and 
$0.31 \times 10^{-7}$ for the halo and disk contribution,
respectively). 

The MACHO team found two microlensing events towards the SMC,
one being a binary event
(Alcock et al. \cite{Alc97c}, Pa\-lan\-que-Delabrouille et al. \cite{Pal97},
Udalski et al. \cite{Uda97}). 
Using a simple estimate for the total exposure and assuming for $\epsilon$ 
the 
corresponding LMC efficiency, they estimate the optical depth towards the 
SMC to be roughly equal to the optical depth towards the LMC.
A similar conclusion has been also reached by the EROS II team, which quotes
a value $\tau \simeq 3.3 \times 10^{-7}$ (Perdereau 1998).
To get a ratio $\tau_{LMC}/\tau_{SMC}$ of about
1, a halo flattening of almost $q_H=0.5$ is required. 
However, it might well be that the events found so far are due to lenses
in the SMC itself (in particular this is the case for the binary event), 
in which case the above conclusions on $\tau$ are no longer valid. 

\begin{figure}
\leavevmode
    \epsfig{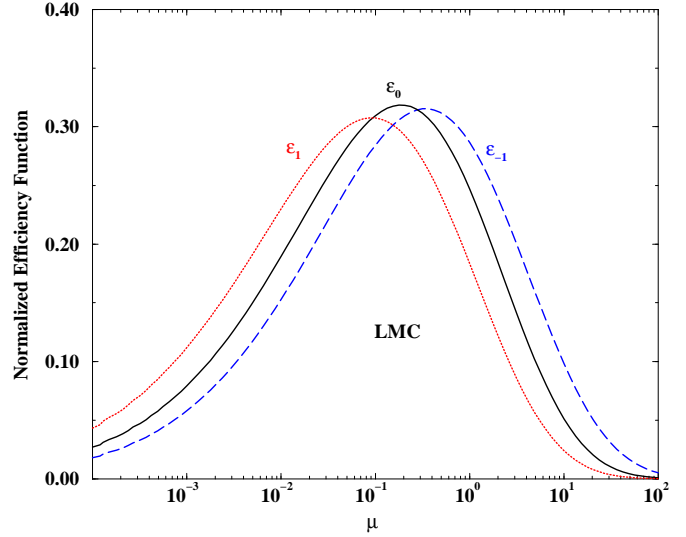}
    \caption{\small{Efficiency functions $\epsilon_{n}(\mu)~(n=-1,0,1)$ 
derived from the sampling efficiency towards the LMC
of the MACHO collaboration.}}
    \label{fig:epsh}
\end{figure}

From the 6 events (excluding the 
binary lens event and an event which is considered being only
marginally consistent with microlensing) 
published by the MACHO team for their first two years observations, 
we find with the method of mass moments for a spherical halo model 
an average mass of 0.26 $M_{\odot}$ (Jetzer \cite{jj}). 
If we assume for the dispersion velocity 191 km s$^{-1}$
as found in the N-body simulation instead of 210 km s$^{-1}$ as used for 
the spherical halo, we get for the average mass a slightly lower value 
of 0.22 $M_{\odot}$. 

Adopting the values for the first two years observation of the MACHO team
towards the LMC,
namely a total exposure of 2.1 years for $8.5 \times 10^6$ stars 
with the detection threshold $u_{TH}=0.661$ and 
the calculated efficiency 
functions $\epsilon_{n}(\mu)$ from the MACHO sampling efficiency,
we compute the expected number of events and the mean event duration
for a $q_H = 0.8$ flattened halo. A delta mass function with $\mu_{H}=0.22$
is used for the halo and the computed power-law mass function
($\mu_{min}=0.021,~\alpha=2.0$) for the disk. We find 
$N_{ev}=12.2$ (where the halo contributes with
$N_{H}=10.4$ and the disk with $N_{D}=1.8$) and $<T>_{LMC}=32.5$ 
days ($<T>_{H}=34.1$ days and $<T>_{D}=23.1$ days, respectively). Using
a spherical halo with $v_{H}=210$ km s$^{-1}$ we find $N_{ev}=14.5$ and 
$<T>_{LMC}=30.3$
days. We see that the flattening of the halo and the change in the
velocity dispersion reduces, although slightly, the number of expected
events, while the average duration increases. 

The value for $N_{ev}$ is valid assuming a halo made entirely of MACHOs.
For a 50\% contribution of MACHOs as implied from
the optical depth we thus expect about 6 events, which compares
well with the 6 observed events. The event duration has to be compared with 
the measured 35.5 days of the 6 observed events. It is noticeable that
we expect about two events as due to the disk contribution.
In addition LMC self-lensing may also contribute with some 
events (Salati et al. \cite{Sal99}), which we do not take into account in our 
calculation. Indeed, the fraction of the optical depth 
due to self-lensing is still controversial,
although a recent analysis of Gyuk et al. (\cite{gdg}) comes to
the conclusion that it contributes at most 10-20\% to the
observed optical depth. Thus the bulk of the events should still
come from MACHOs located in the halo.
Given this mixing of different populations in the events one has to be
careful with the interpretation of the average mass, since it
is derived under the assumption that all events are due
to MACHOs in the halo, which is probably not the case.

\subsection{Dwarf galaxies}

The LMC and the SMC are the only two targets in the halo with a sufficiently 
high number of stars which can be resolved such that 
the method of microlensing can be applied.
Other targets such as the satellite dwarf galaxies, 
can still be used for microlensing observations however, due to their size and
distance much less stars can be resolved.
One could envisage in this case to use the pixellensing 
method instead.
With this method one can look at rich fields of stars which are
located further out and thus are not resolvable.
The method, proposed by Crotts (\cite{Cro92}) and Baillon et al. 
(\cite{Bai92}), has proven to work with M31 and so it is conceavable to 
extend the observations
to some of the several satellite dwarf galaxies 
of the Milky Way.
Such observations would be very useful, since they would allow to test 
other directions as the ones towards the LMC and the SMC.

In 
Table \ref{tab:dwarf} we report the relevant quantities for some possible  
targets in the halo and the neighbourhood of the Milky Way, for which
we assumed 
an extension of the halo of 150 kpc (Bahcall \cite{Bah96})
and a flattening parameter of $q_H=0.8$. 
For the number of events and the mean 
event duration a delta distribution with the mass value $\mu_0=1$ is 
assumed. The dispersion velocity is taken from the simulation 
presented in the Appendix. 
The total exposure is taken to be $10^6$ star-years and the efficiency 
is set equal to 1. The number of events is determined under the assumption
of a halo made entirely of MACHOs.

In comparison for
a standard spherical halo with $v_{H}=210$ km s$^{-1}$ we would
expect about 20\% more events, but with a duration of 
about 7\% shorter. Similar results have also been found
by De Paolis et al. (\cite{Pao96}) in the framework of halo models
with anisotropy in the velocity space.

Other possible targets from which interesting
Galactic structure can be extracted are globular clusters, for which we refer
to the papers by
Gyuk \& Holder (1998), Rhoads \& Malhotra (1998)
and  Jetzer et al. (1998).

\begin{table*}[h!tbp]
 \renewcommand{\arraystretch}{1.2}
  \centering
    \begin{tabular}{|c|c|c|c|c|c|c|c|c|c|c|}\hline
     target& distance &l    &b    & $\tau_{tot}$ &disp. vel&$N_{ev}$&$<T>$
           &$\gamma (0)$  &$\gamma (1/2)$&$\gamma (1)$ \\ 
           & \small{(kpc)} &    &    & \small{$(\times 10^{7})$}&\small{(km s$^{-1}$)}&   
&\small{(days)}&\small{$(\times 10^{2})$}&\small{$(\times 10^{2})$}&\small{$(\times 10^{2})$} \\ \hline
      LMC       &50   & 280.5   & -32.9 &4.8 &191&1.5 &73&28.48&8.60&3.71  \\ \hline
      SMC       & 63  & 302.8   &-44.3  &6.1 &195&1.8 &78&25.69&7.34&3.04  \\ \hline
      NGC 205   &690  & 120.7   & -21.1 &7.3&344&2.8&59 &6.82&1.64&0.62  \\ \hline
      NGC 147   &690  &  119.8     &-14.3  &7.8&195 &1.9 &96&8.02&1.81&0.65 \\ \hline
      Fornax    &230  & 237.3   &-65.7  &6.2  &195&1.4  & 102&7.72&1.65&0.57 \\ \hline
      Draco     &60   &86.4  &34.7  &5.1  &192&1.5  & 77&22.73&6.54&2.74  \\ \hline
      Ursa Minor&80   &104.9  &44.9  &5.0  &184&1.3  &89 &15.06&3.95&1.57 \\ \hline
      Ursa Major&120  &202.3  &71.8  & 5.3 &175&1.2 &106 &9.21&2.10&0.76  \\ \hline
      Leo I     &230  &226.0  &49.1  &6.2  &174&1.3  &115 &7.38&1.60&0.56 \\ \hline
      Leo II    &230  &220.1  &67.2  &6.0  &176&1.2  &113 &7.48&1.59&0.55  \\ \hline
      NGC 2419  &60   &180.4  &25.3  &3.4  &202&1.1  &70 &15.53&4.33&1.80 \\ \hline
      Sagittarius&24  &5      &-15  &12.6  &205&5.2  &56 &143.29&53.74&26.92\\ \hline
    \end{tabular}
    \caption{\small{Optical depth, number of events and event duration
for different targets in the halo and the neighbourhood of the Milky Way. 
For the number of events and the mean event duration a delta distribution 
with the mass value $\mu_0=1$ is assumed. The total exposure is taken to be 
$10^6$ star-years. The halo is assumed to be flattened with $q_{H}=0.8$
and the values for the dispersion velocity are taken from the N-body 
simulation.}}
    \label{tab:dwarf}
\end{table*}

\section{Discussion}

We studied in detail microlensing towards different lines of sight which
are promising to obtain information about the structure of our Galaxy using
the method of mass moments.
With the first year MACHO data in the direction of the Galactic centre
we determined the mass functions of lenses in the bulge and the disk
assuming a power-law.
The obtained mass functions are both slightly less steep 
continuations of the Salpeter IMF down to the brown dwarf region,
with a minimal mass of $\simeq 0.012~ M_{\odot}$ for the bulge and
$\simeq 0.021~ M_{\odot}$ for the disk with 
a slope $\alpha\simeq$ 2.0 for both cases. By varying the 
tilt angle
of the bulge in the range from $15^{\circ}$ to $30^{\circ}$, which is 
suggested by
the most recent models, the inferred mass functions do
practically not change.

As next we computed the expected number of events and their
duration for different targets in the spiral arms, which have
been explored by the 
EROS II collaboration. 
Assuming the same mass function as derived for the disk
we get event durations which are in good agreement with the values
found for the 3 events observed towards $\gamma$ Nor and 
$\gamma$ Sct. On the other hand, the calculated number of events 
turns out to be somewhat smaller, although given the few events
found so far the uncertainties are still large and, moreover, the
theoretical values vary in a significant way
even for slight changes of the parameters such as the velocity
dispersion.

Microlensing towards fields in the spiral arms is 
an important tool to explore the structure of the disk and especially its
mass function and dispersion velocity.
The method of mass moments can be used to get in a systematic way
important information from the 
data, as soon as a sufficient
number of events will be observed, which is certainly the case
towards Baade's Window. It is also important to further develop
the Galactic models, especially the modelling of the
spiral arms and the motion of the stars inside them. Comparison then
with the data will lead to stringent constraints on the model.

To obtain a qualitative understanding how the presence of massive companions
may influence the Galactic halo, we performed a N-body simulation of the
halo from which we obtained a halo flattening parameter $q_H\simeq 0.8$ as
well as an idea how the dispersion velocity can vary as a function of the
observational direction. The parameters from the simulation were then used
to compute the optical depth, the expected number 
of events and the duration for several targets in the halo of our Galaxy.

Although most of the results we obtained have to be considered as preliminary
and will soon improve due to many new events which will be available,
we find that within the adopted model for the Galaxy the agreement
between the computed values for the optical depth, the expected
number of microlensing events and the observation is quite good,
given also the uncertainties in various important parameters.

{\em Acknowledegments.} L.G. and M.S. are partially supported by the Swiss 
National Science Foundation. We thank M. Moniez and R. De\,Ritis for useful 
discussions. Moreover we are grateful to the referee C. Alard for
several useful comments, which helped to improve the paper. 

\appendix
\section{N-body Simulation of the Galactic Halo}


In this Appendix we present a N-body simulation of the galactic halo. 
The simulation aims to give a qualitative understanding of how and to what 
extent 
the presence of massive bodies like the LMC and SMC
influence the shape and the velocity dispersion in the halo.
These latter quantities are relevant for microlensing.
We stress that our simulation has to be considered as a first step 
and that the results we find 
should be regarded as illustrative.

\subsection{The Model}
Our model consists of three massive point particles, which 
represent the Galaxy, the LMC and the SMC, 
respectively. The central mass is assumed to be 
$5\times 10^{11}\,M_{\odot}$, while we set for the LMC-mass
$6\times 10^{9}\,M_{\odot}$ and for the SMC-mass 
$1.5\times 10^{9}\,M_{\odot}$ (Westerlund \cite{westerlund}). In the field 
of these three gravitating bodies we let evolve a halo consisting of 
$2.8\times 10^7$ test-particles for about $9.5\times 10^9\,$years. 
The time resolution of the simulation is $30\,000$ time steps of $10^{13}\,$s
($3.17\times 10^5\,$years) each. The halo is discretised into $50^3$ 
equidistant bins with respect to spherical coordinates
$(r,\,\vartheta,\,\varphi)$, hence, yielding an average population
of $\sim 200$ MACHOs per bin. To keep the problem clean and simple, we
do not yet include an extended (oblate or prolate) massive halo into the 
gravitational potential hence, we ignore any possible backreaction of the halo 
on the clouds. As a consequence of our test-particle approach all the results
will be valid only up to a normalization constant depending on the total
mass of the halo.

The initial conditions for the Magellanic Clouds are chosen such that
at the end of the simulation the locations and the velocities are close
to the observed values  
of the LMC and SMC (Lin, Jones \& Klemola \cite{Lin}). However, since 
the many-body problem depends crucially
on the initial values and we made some compromises to avoid pathological 
orbits due to close encounters between the LMC and the SMC, the final 
location does not exactly match the real one.

The initial spatial distribution of the halo objects mimics an
isothermal sphere with a cut-off radius of $100\,$kpc.
The initial velocities were chosen to be Keplerian in magnitude
and random in direction. We assume a total zero angular
momentum halo in order to not overestimate the halo flattening due to
the presence of the Magellanic Clouds.

To test the reliability of the results we performed
some control runs, varying the initial conditions as well as the duration
and the time resolution of the simulation. 
We found that the results are stable 
unless significant anisotropies are introduced in the initial data. 
A time duration slightly less than a Hubble time is sufficient for the 
developement of a steady configuration.

\subsection{The Results}

The simulation yields the entire phase-space of all halo objects. Hence,
we are able to extract all halo parameters entering in gravitational 
microlensing, which are independent of the sources in relative units with
the exception of the lens mass function. 

As it can be seen in Fig. A1, the final density profile is rather well 
described by an isothermal spheroid
\begin{equation}
\rho(R,\,z) = \frac{\tilde\rho_{\circ H}}{R^2+z^2/q_H^2}
\end{equation}
with $q_H\simeq 0.8$. 
\begin{figure}
\leavevmode
    \epsfig{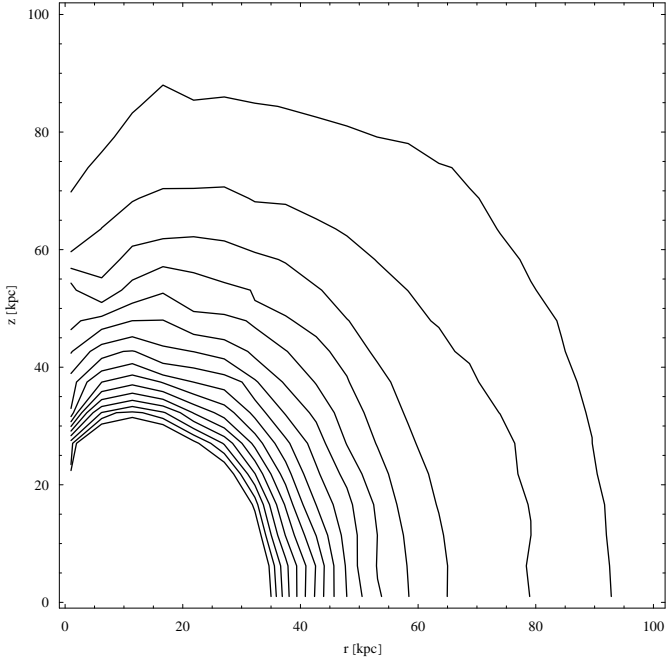}\label{iso}
    \caption{Isocontour lines of the MACHO-density in cylindrical
             coordinates. The final configuration is sligthly flattened
             with $q_H\sim 0.8$. Polar orbits are significantly underpopulated.
}
\end{figure}
An anisotropy in the disk plane is not seen, which is
not surprising due to the model assumptions. The density profile is 
slightly dimpled at the poles which again is a consequence of the LMC and
SMC orbits. The radially integrated MACHO-density decreases towards the
poles and falls below $80\%$ of its average value at a galactic latitude higher than
$\simeq 70\,$ degrees. Although the MACHO-density also varies azimuthally by
about $10\%$, a clear correlation or anti-correlation with the final position 
of the LMC or SMC is not seen.
\begin{figure}
\leavevmode
    \epsfig{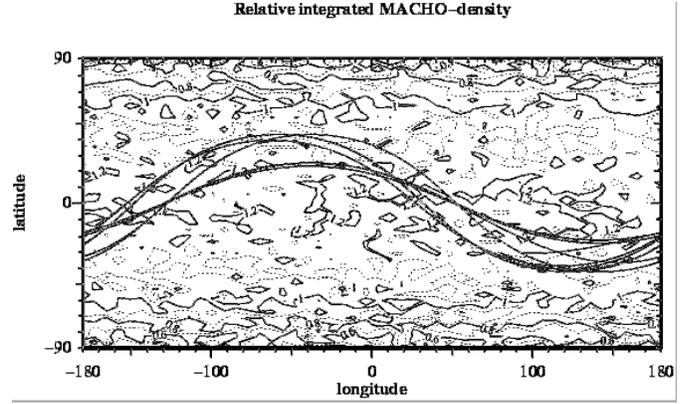}\label{rho_0_50}
    \caption{Relative integrated MACHO-density $\rho$ for an observer in the
             center of the halo. The distance to the source is set
             to be $50\,$kpc. The trajectories show the projected orbits
             of the two massive companions. At the end of the simulation
             the LMC is located at $l=80.7,\,b=\llap{-}31.3$ and the SMC at
             $l=93.7,\,b=\llap{-}30.4$.
}
\end{figure}    
\begin{figure}
\leavevmode
    \epsfig{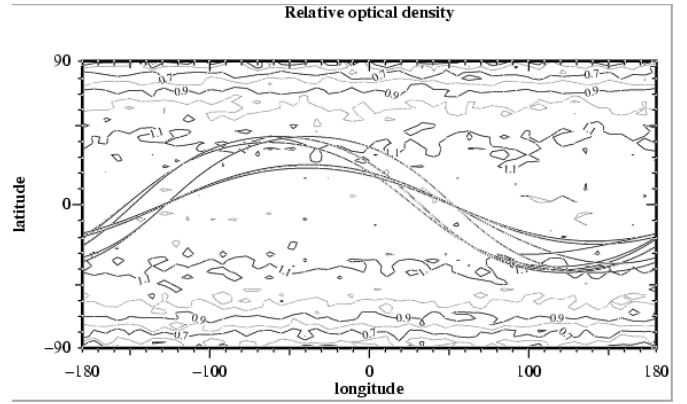}\label{tau_0_50}
    \caption{Relative optical density $\tau$ for an observer in the
             center of the halo. The distance to the source is set
             to be $50\,$kpc. The trajectories show the projected orbits
             of the two massive companions. At the end of the simulation
             the LMC is located at $l=80.7,\,b=\llap{-}31.3$ and the SMC at
             $l=93.7,\,b=\llap{-}30.4$.}
\end{figure}

The angular dependence of the optical density $\tau$ is quite similar to
the radially integrated MACHO-density. Due to the geometrical factor when
integrating along the line of sight, the variation is somewhat weaker. In
fact, the simulation implies that the optical density towards the LMC
should not differ more than $\sim 20\%$ from the average value as a
consequence of the halo flattening induced by the presence of the LMC and
SMC.

Contrary to the density profile the velocity dispersion of the halo objects 
significantly varies with its position in the halo. Hence, the global
distribution function in phase-space cannot be written as a product of 
a spatial and a velocity distribution. In fact, close to the LMC we 
even observe a bulk motion of MACHOs 
which form a separate halo around the LMC. 
Towards an undisturbed direction far from the LMC or the SMC, 
the velocity distribution is rather well 
described by a Maxwell distribution with a dispersion velocity depending on 
the observational position. 

For various locations in the halo the tangential velocity dispersion $v_H$ 
as given by the simulation is tabulated in Table~\ref{tab:dwarf}. 
To calculate the
tangential velocity dispersion, as used in gravitational microlensing
from the simulated spatial one, we use
\begin{eqnarray}
& & v_H^2 = \nonumber\\
& & \sqrt{(v_yv_z\cos\varphi\sin\vartheta)^2+(v_xv_z\sin\varphi\sin\vartheta)^2+(v_xv_y\cos\vartheta)^2}\,,\nonumber\\
\end{eqnarray}
where $\varphi$ and $\vartheta$ denote the azimuthal and the declination
angle of the observational direction in standard spherical coordinates.

\end{document}